# Specifying Autonomy in the Internet of Things:
# The Autonomy Model and Notation

Christian Janiesch
Julius-Maximilians-Universität Würzburg, Josef-Stangl-Platz 2, 97070 Würzburg
Tel.: +49-931-3184930
Fax: +49-931-31849300
E-mail: christian.janiesch@uni-wuerzburg.de
ORCHID: 0000-0002-8050-123X

Marcus Fischer
Julius-Maximilians-Universität Würzburg, Josef-Stangl-Platz 2, 97070 Würzburg
E-mail: marcus.fischer@uni-wuerzburg.de

Axel Winkelmann
Julius-Maximilians-Universität Würzburg, Josef-Stangl-Platz 2, 97070 Würzburg
E-mail: axel.winkelmann@uni-wuerzburg.de

Valentin Nentwich
AUDI AG, 74148 Neckarsulm
E-mail: valentin1.nentwich@audi.de

# Specifying Autonomy in the Internet of Things: The Autonomy Model and Notation

**Abstract**

Driven by digitization in society and industry, automating behavior in an autonomous way substantially alters industrial value chains in the smart service world. As processes are enhanced with sensor and actuator technology, they become digitally interconnected and merge into an Internet of Things (IoT) to form cyber-physical systems (CPS). Using these automated systems, enterprises can improve the performance and quality of their operations. However, currently it is neither feasible nor reasonable to equip any machine with full autonomy when networking with other machines or people. It is necessary to specify rules for machine behavior that also determine an adequate degree of autonomy to realize the potential benefits of the IoT. Yet, there is a lack of methodologies and guidelines to support the design and implementation of machines as explicit autonomous agents such that many designs only consider autonomy implicitly. To address this research gap, we perform a comprehensive literature review to extract 12 requirements for the design of autonomous agents in the IoT. We introduce a set of constitutive characteristics for agents and introduce a classification framework for interactions in multi-agent systems. We integrate our findings by developing a conceptual modeling language consisting of a meta model and a notation that facilitates the specification and design of autonomous agents within the IoT as well as CPS: the Autonomy Model and Notation. We illustrate and discuss the approach and its limitations.

**Keywords**

*Autonomy, agent, Internet of Things, cyber-physical systems, conceptual modeling language, notation*

# 1  Introduction

Digitization changes the characteristics of industries and society sustainably. The emerging Industry 4.0 paradigm for example results in a transformation process that replaces established organization principles and requires novel approaches for the effective design of enterprise operations such as lights-out factories. Further applications range from driver assistance functionality to self-driving cars in the automotive industry and from smart home applications to save energy to smart energy grids for sustainable energy consumption. These scenarios are all characterized by networking between machines and people (Lasi et al. 2014). It is a key challenge for IS research to develop and implement digital artifacts, e.g. methods and techniques, which facilitate the realization of benefits, while accounting for potential risks.

In general, these developments are based on the inter-connection of smart objects or smart devices over a network, for example the Internet, that ultimately merge into an Internet of Things (IoT) (Ashton 2009; Gubbi et al. 2013). Enabled by sensor and actuator technologies, these smart objects increasingly automate industrial value chains forming so-called cyber-physical systems (CPS) to deliver smart services (Geisberger and Broy 2012; National Science Foundation 2016; Beverungen et al. 2017).

A growing number of sensors boosts the generation of large amounts of data, which can help to design more economical systems but also systems that are more environmentally-friendly, energy-efficient, and, thus, more sustainable in general. Research on artificial intelligence and big data analytics have greatly improved the means of analyzing and aggregating this data to meaningful decision-relevant information. Still, many decisions need to be taken, approved, or executed by humans. This situation poses a significant challenge for today's decision-makers as their ability for problem solving is limited by their cognitive capabilities. They cannot address all necessary decisions in a timely manner without passing up optimization potential or making mistakes (Onnasch et al. 2014).

One way of addressing these challenges is to increase the autonomy of machines by equipping them with the necessary capabilities for independent decision making. Consequently, machines become self-determined and self-contained autonomous agents that situationally control their actions. For example, in the software industry, there is a trend to autonomously select and apply software through online updates and only (sometimes) inform the user about the process.

However, research and practice have not yet defined a generally accepted methodology for the design and implementation of autonomy of (artificial) agents in the IoT or in a CPS as it is necessary to define boundaries for this autonomy in a consistent way. Hence, there is a gap between the general idea and implementation of autonomous smart objects in the IoT, which currently is predominantly implicit, and the practical and explicit design of autonomy in such systems. We aim to close this gap.

In the following, we address this gap by answering the following questions:

*(1) How can we conceptualize and structure autonomy in the IoT?*

*(2) What design requirements can be derived from the characteristics of and interactions between autonomous agents in the IoT?*

*(3) What is the appearance of a graphical conceptual modeling language that can be used to model autonomy independent of the concrete implementation of agents?*

These questions also entail, that we do not attempt to create yet another modeling approach to model the structure of complex multi-agent systems or interaction processes but a superset of concepts necessary to model autonomous behavior. This superset can be used to enhance existing modeling languages or on its own for the specific purpose of modeling autonomy. It has practical value as a design language with the ability to specify the complex parameters of autonomous agent interaction, which is currently not possible in common design languages. To our knowledge, it is the first graphical conceptual modeling language that focuses on autonomy in information system design.

This paper is organized as follows: Section 2 introduces our research design. Section 3 reviews necessary theoretical foundations on the IoT, CPS, and autonomous agents. We analyze existing systematizations of IT autonomy and their modeling. Drawing upon these implications, we derive a set of constitutive characteristics and interaction patterns of autonomous agents in the IoT. We use these to define design requirements for their conceptualization in Section 4. We integrate our findings by introducing a comprehensive meta model in Section 5. We transform the meta model into a graphical notation. Both form the *Autonomy Model and Notation (AMN)*, a graphical conceptual modeling language which we demonstrate and evaluate in a smart factory scenario in Section 7. In Section 8, we conclude by summarizing results, limitations, and future research opportunities.

# 2 Research Design

For this research, we apply a problem-centered Design Science Research (DSR) approach as suggested by Peffers et al. (2007) aligned with the seven guidelines by Hevner et al. (2004). Problem-centered DSR provides a nominal process to better structure and conduct research as well as a mental model to communicate its output (Peffers et al. 2007).

Being experts in the design of domain-specific conceptual modeling languages, we notice that despite rich scientific discussion in other fields of research, there was no means to conceptually specify autonomy for services and, thus, for smart objects, smart services, or even smart service systems other than using non-functional attributes in an unstructured and ad-hoc manner. Our contribution addresses this important unsolved problem in a unique and innovative way. Typical outcomes of DSR activities are artifacts, which include constructs, models, methods, and instantiations (March and Smith 1995). On the one hand, we propose major characteristics of autonomous agents and their ways and means of interactions. We do this by integrating implications from other fields of research to formulate design requirements for specifying autonomy in the IoT. On the other hand, we face the unsolved problem of modeling and conceptualizing autonomous behavior in the IoT. We tackle this by developing a graphical conceptual modeling language consisting of two novel artifacts: a meta model and a graphical notation.

We summarize the applied DSR approach as well as complementary methods and resulting artifacts in Figure 1.

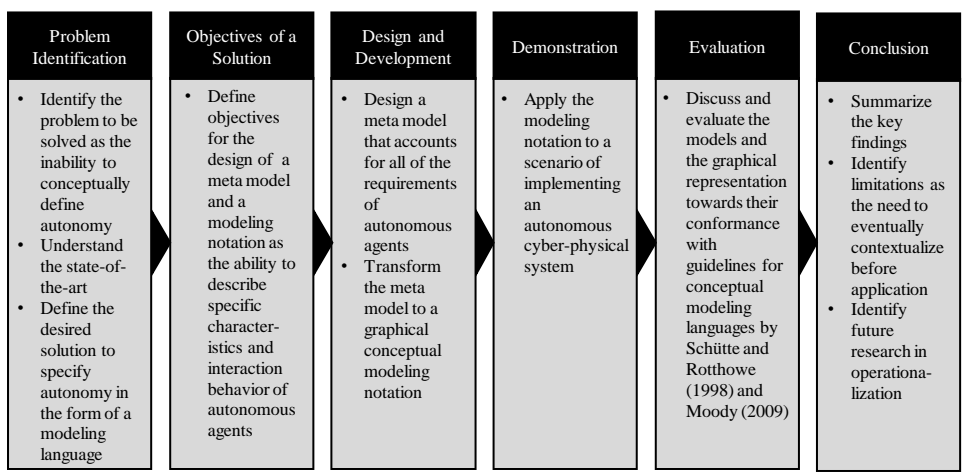

**Figure 1. Research Approach**

Our research comprises six main activities. As our research progresses, we iteratively develop the configuration of the resulting artifacts. During the phase of *problem identification*, we highlight the importance of the problem for research and practice and identify the research gap as the result of the difference between the current state of research for the IoT and CPS and the requirements of the problem that this contribution addresses. We use forward and backward literature search with starting keywords such as "autonomous agent", "machine autonomy", and "conceptual model autonomy". We then define *corresponding objectives of a solution* by deriving design requirements that guide the design of a modeling language's meta model and notation, namely the ability to model characteristics as well as interaction behavior. Our *design and development* activities focus on the development of a comprehensive meta model that satisfies our design requirements by incorporating all relevant approaches from literature. We subsequently transform the meta model into a graphical notation, which supports the specification of autonomy in the IoT and CPS. To *demonstrate* the main artifact of our contribution, we use the scenario technique and apply the proposed notation to a scenario of a CPS in an Industry 4.0 smart factory scenario in the smart service world. Our *evaluation* section is based on multiple techniques common to assess the outcomes of DSR. Initially, we use informed argumentation to build an argument for our artifact's applicability for specifying autonomy in the given scenario and beyond. Then, we compare our model and notation with requirements from conceptual modeling language design and evaluation literature. We further define additional scenario candidates to specify the potential utility of our notation and to widen the scope of its application. During the *conclusion* phase, we highlight the limitations and constraints of our artifact and identify future research potentials.

Following the knowledge contribution framework of Gregor and Hevner (2013), we consider our DSR contribution an adaptation of conceptual modeling research to the domain of agent and autonomous systems for the design of IoT-based systems and CPS. In doing so, we adapt those aspects of the state-of-the-art that are meaningful to the design of conceptual modeling languages providing a new area of application for them. Our contribution has explanatory power as well as it provides design practice theory for the design and implementation of autonomy and multi-agent systems.

# 3 Theoretical Foundations

## 3.1 Internet of Things and Cyber-physical Systems

According to Ashton (2009) and Weiser (1991) and refined by Gubbi et al. (2013), the IoT describes objects, whose capabilities go beyond the autonomous collection of local data. These objects learn from and adapt to the collected data as well as use it to interact with the world around them, while communicating globally over a network, for example, the Internet.

Although CPS have been researched extensively in recent years, neither research nor practice provide a generally accepted definition. Based on the characteristics of the IoT, Gill (2008) describes CPS as IoT-based systems, which are physical, biological, or structurally engineered systems that are integrated, controlled, and operated by a software system. The software system is an embedded and/or distributed system that typically requires real-time response functionalities. Lee (2008) further highlights the interdependencies that exist between embedded computers and physical processes within feedback loops. Geisberger and Broy (2012) describe the necessity for single or multi-modular human machine interfaces that support communication and interaction processes in CPS to facilitate the global applicability of data and services in local operations.

The IoT can be employed in a vast amount of use cases from the management of automated factories, the provisioning of reliable decentralized energy networks, the monitoring of our environment to the coordination of transportation services. For example, in a transportation scenario for a mass event at a stadium the following is conceivable (cf. Figure 1). Traffic in the vicinity of the stadium can be shaped by variable speed limits on smart signs (1), autonomously driving cars can be send to available parking spots (2), intelligent car-sharing vehicles can try to accommodate multiple passengers for similar destinations when vehicles run short (3), and in the event of an emergency, spectators need to be evacuated to a safe distance (4). Many of these tasks can be automated but not all of the tasks should autonomously automate.

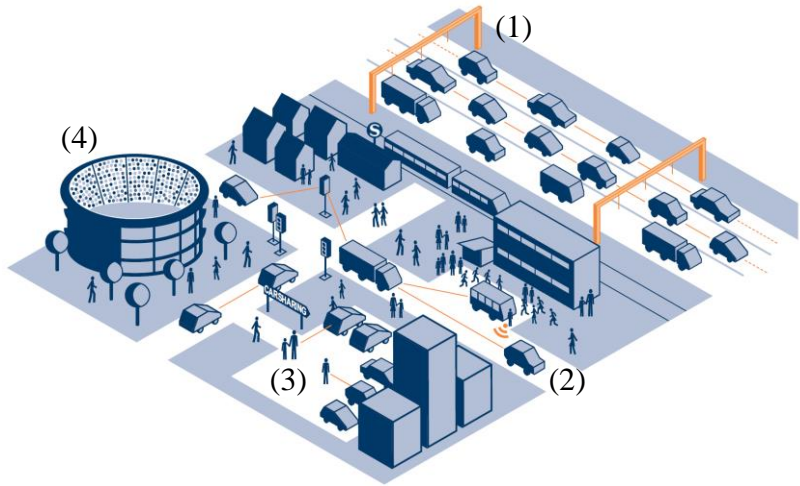

**Figure 2. IoT Scenario (cf. Geisberger and Broy 2012)**

In summary, CPS are IoT-based software-driven systems that integrate mechanic and electronic components, which communicate in real-time over a global network infrastructure, while deploying their physical actions locally. Due to the varying degree of autonomy and automation, implications and requirements emerge for human-computer interaction (Onnasch et al. 2014).

The European Commission has realized the importance of the IoT and established the European Research Cluster on the Internet of Things (IERC) in 2007 to identify technology research challenges at a European and global level. Projects deal with topics such as security, interoperability, business value, and industry application. The National Institute of Standards and Technology (NIST) has also advanced many related activities in particular in the area of cloud and fog computing as well as IT security. While many projects and frameworks consider the interaction of agents and their configuration, we cannot identify a project that specifically researches the specification of autonomy for smart devices.

## 3.2 Agents and (IT) Autonomy

Agents are entities, for example machines but also humans, that collect data about their environment through sensors and use this information to perform actions through actuators (Russell and Norvig 2014). In the context of the IoT, agents can be anything from smart ceiling lights to autonomous vehicles. Autonomous agents pursue their own agenda and independently determine their course of action (Franklin and Graesser 1997). Consequently, an agent's most important components are its interfaces, which are its sensors to observe its environment and its actuators to deploy actions within its sphere of influence. According to Schillo and Fischer (2003), the environment represents a multi-agent system, where each system can be regarded as an agent itself that can be part of another, larger multi-agent system.

In general, autonomy describes an entity's or agent's ability to act independently and self-determined. While self-determination allows for acting on one's own responsibility, independence entails that an agent's actions are not controlled by an external instance (Bradshaw et al. 2004).

In the IS domain, autonomy can be considered as a system's non-functional characteristic, which can be used to describe other functions. It can be regarded as a specialization of adaptivity, summarizing an agent's ability to accomplish its goals based on a set of input parameters, while continuously adapting its behavior to environmental changes.

Automation is a related concept. It enables machines to perform actions without human support or interaction until decisions need to be made. Eliminating the necessity for human interventions in processes and tasks increases efficiency and speed while reducing the need for physical or cognitive human actions (Chestnut 1963). In coordinative functions, IT automation can reduce human responsibilities and improve operational speed and safety. It is a prerequisite for IT autonomy which requires automated analyses for decision-making.

In addition, autonomy does not only apply to individual functional aspects of agents but can be distinguished by different degrees of automation based on prevalent interaction patterns between humans and machines (cf. among others Parasuraman and Riley 1997). The resulting classification defines a continuum that spans from complete agent autonomy on the one end, with machines acting autonomously based on their own goals, to physical and cognitive actions exclusive to humans on the other end (Onnasch et al. 2014) (cf. also Section 3.6).

Thus in summary, IT autonomy describes the ability of an (artificial) agent to make decisions and execute corresponding actions in an independently and self-determined manner.

## 3.3 Types of Autonomous Agents

Although Russel and Norvig (2014) do not address the integration of logic, goals, and moral concepts into an agent's information processing capabilities, they define four archetypes of agents. Table 1 gives an overview.

**Table 1. Types of Agents (Russell and Norvig 2014).**

| Type | Description |
| --- | --- |
| Reflex agents | They perform actions by following a set of event-condition-action-rules. They operate by identifying a rule that matches the condition of the current situations and deploy the action associated with that rule. |
| Internal state agents | They augment the reflex agent with the capability to integrate historical and trend environment information into the decision-making process to determine the best possible course of action. |
| Goal-based agents | They augment the internal state agent with the capability to evaluate available action candidates towards a set of goals that specify desired environmental conditions. Agents are provided with a certain degree of freedom, as their actions are not rule-based but rather aimed towards the accomplishment of predefined goals (Munroe and Luck 2003). |
| Utility-based agents | Augments the goal-based agent with the capabilities to evaluate available action candidates towards a utility function. Agents of this type deploy actions that maximize their expected returns. |

Autonomy is a property independent of the agent type. Yet the processing and decision-making capabilities clearly have an impact the ability to take autonomous decision.

## 3.4 Properties of Autonomous Agents

While human agents primarily use their eyes and ears as sensors and their mouth and hands as actuators, artificial agents comprise a broader range of interfaces, ranging from cameras and thermometers as sensors to robotic components and digital signals as actuators. Both are constrained by (physical) capacity limitations.

To enable the exchange of data across multiple agent systems, an adequate communication medium is required. Agah (2000) introduces six options for communication between humans and machines that are based on the five natural senses of sight, hearing, smell, taste, and touch but also consider biometric data such as body temperature.

Autonomy gives the agents the ability to decide self-determined when to perform which tasks and when and whom to communicate with through a chosen channel. In this way, agents are able to react to foreseen and unforeseen events in their environment, solve expected and unexpected problems and – in general – be part of dynamic systems (Jennings 1999; Munroe and Luck 2003). Franklin and Graesser (1997) summarize properties that can be used to classify agents (see also Table 2). According to them, autonomous agents satisfy at least the first four properties: reactive, autonomous, goal-oriented, and temporally continuous. Reactiveness warrants timely responses to asynchronous, external stimuli and is a key characteristic of autonomous agents. Hence, information processing should take place in real-time (Brustoloni 1991).

Table 2. Properties to Classify Agents (based on Franklin and Graesser 1997)

| Property | Meaning |
| --- | --- |
| Reactive | Responds in a timely fashion to changes in the environment |
| Autonomous | Autonomous exercises control over its own actions |
| Goal-oriented | Does not simply act in response to the environment |
| Temporally continuous | Is a continuously running process |
| Communicative | Communicates with other agents, including people |
| Learning | Changes its behavior based on its previous experience |
| Mobile | Able to transport itself from one machine to another |
| Flexible | Actions are not scripted |
| Character | Believable "personality" and emotional state |

Wooldridge and Jennings (1995) emphasize social characteristics as a complementary property. They focus in particular on the communication with other artificial or human agents. In doing so, they characterize an agent's social self-concept or social rationality. While credibility describes the believability of an agent and, thus, its reliability from the perspective of the receiver, rationality characterizes agents who are agreeable to reason and, thus, act according to a specific social agenda. Kalenka and Jennings (1999) distinguish three forms of social competencies that are relevant for the decision-making process of autonomous agents (see Table 3).

**Table 3. Social Competencies of Agents (Kalenka and Jennings 1999).**

| Type | Description |
|---|---|
| Socially self-interested | An agent of this group concentrates on his goals and benefits. He executes actions accordingly while ensuring not to be overly detrimental to the society. |
| Helpful | An agent of this group choses alternatives that are beneficial for the society but not detrimental to himself. Hence, he might execute actions without personal benefit. |
| Cooperative | An agent of this group might execute actions that are detrimental to itself if they are compensated by other agent's actions in the society that in sum realize a benefit. |

Any decision should be taken reviewing ethical circumstances. While model logic distinguishes what is permitted and what is required, normative decisions may not always be ethical ones. Moor (2006) describes several types of ethical agents and concludes that in the near future it remains unclear whether artificial agents can become full ethical agents. Anderson and Anderson (2011) and Pereira and Saptawijaya (2016) provide further facets to the discussion. See Table 4 for an adapted summary based on Moor's conceptualization.

**Table 4. Ethicality of Agents (based on Moor 2006).**

| Type | Description |
|---|---|
| Non-ethical agent | An agent is of this group when its programming does not consider ethics and it has no further facility to incorporate any aspects of ethics. |
| Implicit ethical agent | An agent is of this group when its programming already addresses ethics such as safety or critical reliability concerns. |
| Explicit ethical agent | An agent is of this group when it has the ability to adjust judgements or prioritize based on ethical considerations. |
| Full ethical agent | An agent is of this group when it can make explicit ethical judgments and generally is competent to reasonably justify them similar to an average adult human. |

Based on these findings, we can state that a machine's degree of autonomy is also influenced by its information processing capabilities and its social and ethical interactivity with others.

## 3.5 Patterns of Autonomous Agent Interaction

Furthermore, the information gradient, which is a result of the ratio of data processed within a system to the data that is exchanged between systems, can be an indicator for a system's degree of autonomy (Richling et al. 2011). To specify an agent's information processing capability, Endsley and Kaber (1999) as well as Parasuraman et al. (2000) draw upon the functions of monitoring, generating, selecting, and implementing. We use these functions to describe the general procedure of information processing: sensory processing, perception, response selection, and response execution (cf. for more detail on information processing Wickens and Hollands 2000). While Endsley and Kaber (1999) derive a system's degree of autonomy by assigning functions to either a human or a machine, Parasuraman et al. (2000) focus on the different autonomy levels within the functions themselves.

Yanco and Drury (2004) provide a taxonomy of eight potential interaction patterns between humans and machines, specifically robots (see Figure 3).

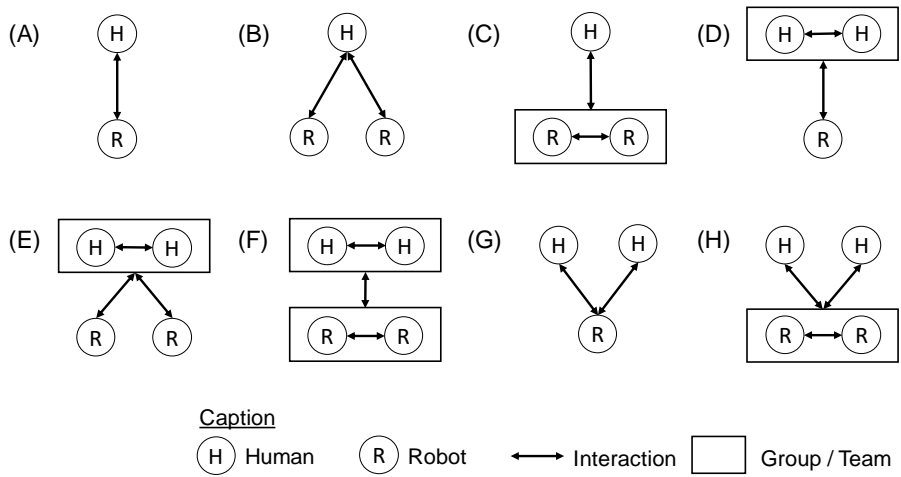

**Figure 3. Interaction Patterns between Humans and Machines (based on Yanco and Drury 2004)**

(A) describes a situation, in which the human agent controls the interaction. The machine focusses on collecting environmental information and forwards it to the human. It also executes the tasks delegated by the human. In situation (B), a human agent controls two or more machines by organizing and coordinating necessary tasks. Subsequently, machines perform these tasks independently. The interaction pattern in (C) describes two or more machines controlled by a human agent that receive instructions from the human but organize, coordinate, and execute the necessary tasks autonomously in a group.

In (D), two or more human agents collaborate in a group to collaboratively define and coordinate tasks. These tasks are transferred to a machine for subsequent execution. In situation (E), two or more human agents collaborate in a group to collaboratively define and coordinate tasks. These tasks are transferred to two or more machines. Each machines executes its assigned tasks independent from other machines. The interaction in (F) is characterized by two or more human agents that collaboratively define and coordinate tasks. These tasks are transferred to two or more machines in a group that coordinates the execution of these tasks autonomously.

In (G), two or more human agents define tasks and transfer them to a machine independently. The machine then coordinates the execution of the incoming tasks autonomously. Ultimately, situation (H) specifies two or more human agents that define tasks and transfer them to multiple machines in a group independently. The machines organize, coordinate, and prioritize the execution of the incoming tasks autonomously. The case, where multiple humans instruct multiple machines independently, is a combination of situations (A), (B), and (G) but not a case of its own.

The capability to form functional groups are similar to the characteristics of complex adaptive systems as described by Holland (2006). These systems comprise groups of agents that interact and learn to adapt.

Finally, Lee and See (2004) argue that mutual trust is fundamental for reliable communication processes. Trust is an important factor if an own evaluation of action candidates is not possible due to uncertainty or complexity. The level of trust states whether a decision will be accepted or if the decision-making process will be performed independently. To measure the trust in communication processes, one can use the attributes of reliability and conformity (Parasuraman and Wickens 2008; Wickens and Hollands 2000). Furthermore, the security of the communication has to be considered. Following Dresner and Stone (2008) for the case of autonomous driving, we focus primarily on the possibility of communication channel corruption. Nevertheless, at least privacy, safety, and trustworthiness of agents and communication channels can also impact behavior. For the context of this paper, we assume credible communication without outside attempts to influence or harm the involved agents. Ren et al. (2014) and Zhou et al.

(2018) provide starting points for IoT security independent of autonomous behavior.

## 3.6 Levels of Agent Autonomy

According to Sheridan et al. (1978), levels of automation and autonomous decision-making can be distinguished on a ten-staged scale. With complete autonomy to both ends of the scale, they describe more differentiated interactions of humans and machines on the levels in between.

In line with that, Parasuraman et al. (2000) also use ten levels of autonomy. Each level can be applied to the phases of information processing of humans. In addition to the state of complete autonomy, the concept differentiates each level by considering the type of information, power of veto, and the number of offered action candidates. Sheridan and Parasuraman (2005) reduce the scale to eight stages when examining the case of computer-sided action selection.

Sheridan (1997) introduces a concept that considers the fundamental characteristics of supervisory control. It distinguishes different automation levels on a five-staged scale and draws upon the processing of information. Sheridan (2011) further defines seven types of influence factors that alter an relationship between autonomous agents by changing system parameters. Consequently, the authors widen the scope of current approaches by reflecting that autonomy does not exclusively result from one's own authority during decision making, but it is also affected by other entities within the same system.

Castelfranchi (2000) proposes 13 dimensions that determine the degree to which an agent is dependent on its environment. Schillo and Fischer (2003) use these characteristics to investigate autonomy in multi-agent structures.

Endsley (1987) distinguishes four levels of automation. Endsley and Kiris (1995) use this distinction and specify five levels of control: no system support, decision support, consensual artificial intelligence, monitored artificial intelligence, and full automation. Endsley and Kaber (1999) further extend the schema into a ten-staged taxonomy, ranging from manual control to full autonomy.

Jipp (2014) also classifies different levels of automation (1) monitoring and reasoning, (2) modelling, path planning, and navigation, (3) collision avoidance, and (4) low-levels of control. In line with other concepts, she integrates different autonomy levels that range from complete control by humans to humans exclusively monitoring the target achievement process.

Kalenka and Jennings (1999) examine the social self-concept of agents within groups and associate autonomy with an agent's social behavior that results from its goals and its preference toward the maximization of individual or group benefits. In line with that, Bradshaw et al. (2004) adjust autonomy by examining agents that adapt their autonomous behavior to changes in their permissions, possibilities, capabilities, as well as to obligations to related agents that jointly perform tasks to accomplish a set of shared objectives.

Finally, Onnasch et al. (2014) analyze the impact of different autonomy levels on the productivity of humans that are affected by automation. The authors argue that there is a trade-off between benefits and costs. Despite the improved execution performance of routine tasks, even in scenarios with varying loads, the study indicates that automation decreases performance and situational awareness when decision support is not working properly.

As outlined above, several contributions distinguish different interaction levels between human and artificial agents to classify autonomy. We integrate and summarize these levels in Table 5. The table is somewhat ordered in terms of decreasing autonomy. However, it is not possible to arrange the following 20 levels in a distinct order representing full machine autonomy (M, level 1) to full human autonomy (H, level 20).

**Table 5. Summarizing View on Levels of Autonomy**

| No. | Description | Reference |
|---|---|---|
| 1 | M performs decision-making and corresponding actions autonomously, H is not involved | (Endsley 1987; Endsley and Kaber 1999; Onnasch et al. 2014; Parasuraman et al. 2000; Sheridan and Parasuraman 2005; Sheridan and Verplank 1978) |
| 2 | M performs decision-making and corresponding actions autonomously, while H monitors certain aspects of M and intervenes if necessary | (Sheridan 2011) |
| 3 | M performs decision-making and corresponding actions autonomously, while H monitors M and intervenes if necessary | (Endsley 1987; Endsley and Kaber 1999; Jipp 2014) |
| 4 | M performs decision-making and corresponding actions and randomly delegates a case to H | (Endsley and Kiris 1995; Onnasch et al. 2014) |
| 5 | M performs decision-making and corresponding actions until H vetoes against them | (Onnasch et al. 2014) |

| 6 | M performs decision-making and provides H with the opportunity to veto against it for a limited time | (Parasuraman et al. 2000; Sheridan and Parasuraman 2005; Sheridan and Verplank 1978) |
|---|---|---|
| 7 | M performs decision-making and executes the corresponding tasks only after H's approval | (Endsley 1987; Endsley and Kaber 1999; Parasuraman et al. 2000; Sheridan and Parasuraman 2005; Sheridan and Verplank 1978) |
| 8 | M provides H with a single action candidate that H can refuse | (Endsley 1987; Onnasch et al. 2014; Parasuraman et al. 2000; Sheridan and Parasuraman 2005; Sheridan and Verplank 1978) |
| 9 | M provides H with a selection of action candidates | (Endsley 1987; Onnasch et al. 2014; Parasuraman et al. 2000; Sheridan and Parasuraman 2005; Sheridan and Verplank 1978) |
| 10 | M provides H with all action candidates | (Parasuraman et al. 2000) |
| 11 | M and H collaboratively define action candidates, while H performs decision-making | (Endsley and Kaber 1999) |
| 12 | H provides M with action candidates, while M performs decision-making and executes corresponding actions | (Endsley and Kaber 1999) |
| 13 | M performs actions and notifies H at its own discretion | (Parasuraman et al. 2000; Sheridan and Verplank 1978) |
| 14 | M performs actions and notifies upon H's request | (Parasuraman et al. 2000; Sheridan and Parasuraman 2005; Sheridan and Verplank 1978) |
| 15 | M performs actions und notifies H at all times | (Parasuraman et al. 2000; Sheridan and Parasuraman 2005) |
| 16 | H performs decision making and corresponding actions autonomously until M takes over to execute the remaining tasks autonomously | (Endsley and Kaber 1999; Sheridan and Verplank 1978) |
| 17 | H performs decision making and transfers corresponding tasks to M for execution | (Endsley and Kaber 1999) |
| 18 | H performs decision making and controls M | (Jipp 2014) |
| 19 | H performs decision-making and controls M, but M is partially self-controlled and executes sub-tasks autonomously | (Jipp 2014) |
| 20 | H performs decision-making and corresponding actions, while M is not involved | (Endsley 1987; Endsley and Kaber 1999; Parasuraman et al. 2000; Sheridan and Parasuraman 2005) |

These levels provide input on items necessary to make or delegate decisions as well as interactions patterns between agents which should be part of any procedural specification of agent behavior. In the context of this paper, we do not further discuss autonomy interaction patterns.

### 3.7 Conceptual Modeling of Agents and IT Autonomy

To the best of our knowledge, there is no approach to graphically model aspects of autonomy other than through textual annotation or attribution. While there are several general-purpose conceptual modeling languages to model the structure and behavior of systems, none of them considers autonomy as more than a functional requirement.

Considering the structure of the systems, the entity-relationship (ER) model (Chen 1976) covers the data perspective and allows for the definition of (data) entities and their relations. It may be used to structure conceptual components and their relations of agents, but it does not include constructs to model autonomy. The Unified Modeling Language (UML) (Jacobson et al. 1999) includes components as well as class diagrams to model structure and activity diagrams as well as interaction diagrams to model behavior. Again, none of the above consider autonomy constructs.

In computer science, several design methodologies for multi-agent systems are being discussed including AML, AOR, AUML, Gaia, MAS-ML, MESSAGE, or Tropos (Bresciani et al. 2004; Odell et al. 2000; Evans et al. 2001; Wooldridge et al. 2000; Silva et al. 2004; Wagner 2003; Cervenka and Trencansky 2000). However, none of them is generally accepted or explicitly includes autonomy.

The Business Process Model and Notation (BPMN) provides a comprehensive standard to model behavior and interaction of processes (Object Management Group 2013). While automation and event-condition-action are properties of the modeling languages, autonomy is not.

Practice-oriented specifications such as the Fundamental Modeling Concepts (FMC) (Apfelbacher and Rozinat 2003) and its extension, the Technical Architecture Modeling (TAM), or ArchiMate (The Open Group 2017) reuse some of these concepts but do not extend them to cover autonomy.

There are domain-specific modeling languages for IoT-based smart service systems. These range from technical to business-oriented languages (cf. e.g. DNN, e3, OSSR, OSSN, USDL, SNN) (Barros and Oberle 2012; Bitsaki et al. 2008,

Scholten, 2013 #1532; Cardoso 2013; Cardoso et al. 2013; Kartseva et al. 2010; Danylevych et al. 2010). They focus primarily on service delivery, interoperability, and business models. Autonomy is not considered either.

## 4 Design Requirements for Modeling Autonomous Agents

### 4.1 Characteristics of Autonomous Agents

In the following, we provide a summary of the previous section as well as a derivation of design requirements for the modeling of autonomy. Agents always act in an environment. Consequently, agents must collect data through sensors, use this data to select adequate actions, and act upon their environment through actuators. We define an agent's environment as a multi-agent system. At a high level, one can observe how an agent interacts through its sensors and actuators. An agent, which appears as a single entity to the outside world, may at a lower level, be composed of many sub-agents and vice versa. They have a recursive relationship that allows for an arbitrary level of granularity. Further, we need to distinguish between (abstract) types of agents that can be instantiated to ease grouping. Each of these may have different autonomy characteristics. The necessity for modeling complex multi-agent structures results in the first design requirement (DR):

> *DR 1: The language must support abstract and concrete, hierarchical structures of autonomous agents.*

Based on different configurations, an agent can be equipped with an arbitrary number of sensors and actuators. While actuators transmit information of a certain type, this data is collected with compatible sensors.

> *DR 2: The language must provide constructs to distinguish different types of sensors and actuators and specify which of them are supported by agents. Furthermore, it must specify the compatibility of these interfaces.*

The direction of an interaction indicates the flow of information and reveals prevalent dependencies between agents and the superstructure of a multi-agent system.

> *DR 3: The language must provide constructs that define an interaction's direction to reveal its inherent communication structure.*

We consider an agent as autonomous if it can perform tasks independently and self-determined. This requires autonomous decision making concerning the allocation of resources and the scheduling of task execution.

> *DR 4: The language must provide a set of simple and complex rules to specify an agent's behavior and decision-making process.*

Independent decision making, as an essential part of an agent's autonomy, must take into consideration the internal and external states and follow certain criteria that guide the evaluation of action candidates.

> *DR 5: The language must provide constructs to specify different goals, utility functions, and states.*

Analyzing large quantities of sensor data is restricted by the requirement to respond to stimuli in real time. In particular, humans have a limited capacity of attention. It limits a human's short-term working memory as well as its ability to collect and process information, to select from multiple action candidates, and to execute the selected actions respectively. Due to economical or technological constraints, these limitations can also apply to artificial agents.

> *DR 6: The language must provide constructs that specify functional capacity restrictions, similar to the concept of human attention.*

Multiple agents can form a group or a team. For example, a human can assign a task to a group of machines that coordinates and executes these tasks independently. As specified by DR 1, each agent can comprise an arbitrary number of sub-agents, which can result in conflicting goals, utility functions, and states. They need to be consistent.

> *DR 7: The language must incorporate the concept of social and ethical rationality. It must specify different types of social and ethical self-concepts and allow to group agents accordingly. This also applies to agents of the same type if their communication behavior is determined explicitly. Thus, the language must provide constructs to specify agent classes that differ in their number of potential and concrete instances.*

In addition, it is important to note as to whether agents believe information they receive and if their actions conform to the expectations. Adequate security mechanisms can further prevent corruption during the communication processes. Those mechanisms provide means to increase the reliability of the communication itself. While there is a large body of research on the safety, privacy, and trustworthiness of communication, we abstract from concrete mechanisms and summarize the requirement as:

> *DR 8: The language must allow the specification of an agent's confidence in the communication channel and provide constructs for reliability, conformity, and security.*

The concepts of coordination protocols and supervisory control (Sheridan and Verplank 1978) suggest that agents' states and behavioral rules provide adequate structural specification for their coordination. Consequently, we can describe an agent's group behavior by specifying its states and behavioral rules. As this is accounted for by DR 5, coordinating group behavior does not impose further design requirements.

## 4.2 Interaction of Autonomous Agents

During our analysis of related work, we summarized the related work on autonomous agent interaction into 20 levels of autonomy (cf. Table 5).

The two extreme cases on level 1 and 20 represent a situation where either the machine or the human has full control over decision-making and task execution. DR 1 to 8 specify autonomous agents suffice to model these two cases. These two levels do not result in further design requirements.

In level 2 and 3, one agent performs decision making and executes the corresponding tasks, while the other agent is a controlling instance that intervenes if necessary. We distinguish both levels by the degree of supervision that results from the number of controlled operations and functionalities. Both levels conform to the concept of supervisory control and its requirements respectively.

Level 4 tackles the out-of-loop performance problem to avoid loss of skills and of situational awareness as information processing shifts from active to passive for the human agent.

By introducing the concept of interaction between autonomous agents, we defined a set of design requirements for specifying the exchange of data objects between two or more agents. The focus here is on the exchange of tasks and subtasks, which is a new design requirement.

It is important to note that we specify generic design requirements independent of specific use cases. Hence, to distinguish for example autonomous replies to maintenance requests but non-autonomous behavior in other cases, one can use said data objects in any applicable level of automation.

*DR 9: The language must support task lists as a special type of data objects.*

Levels 5 to 7 specify situations, in which machines perform decision making autonomously and execute corresponding tasks. Both levels differ by the extent to which a human can veto against a decision and, thus, interrupt or impede the execution of a certain task.

*DR 10: The language must provide data objects that indicate whether an agent accepts or declines a decision and orders to execute certain tasks.*

A similar situation to level 4 arises in levels 8 to 10, which entail that human agents are provided with a varying number of action candidates, ranging from (1) all candidates and (2) a selection of candidates to (3) a single option. Similarly, the levels 11 and 12 describe an identical situation but with a reverse flow of information.

*DR 11: The language must provide data objects of the type task list corresponding to the scope of action candidates.*

Levels 13 to 15 describe situations, in which machines are responsible for the execution of tasks and use different communication behavior to notify other agents about their state. Agents send notifications (1) at their own discretion, (2) only in response to request by other agents, or (3) at all times.

*DR 12: The language must provide constructs to specify the type of a notification.*

We can further group levels 16 and 17, as they constitute situations in which a human agent performs decision making and transfers tasks. Both levels differ by the degree to which the tasks are split between the different agents. The requirement of sending tasks is already covered by DR 9.

Similar to the levels 2 and 3, levels 18 and 19 describe human agents either interacting as passive observers that only intervene if necessary or they take an active role by participating in the decision-making process and by controlling other agents. We distinguish both levels by the frequency and extent of human interventions that is correlated with the complexity and importance of the underlying tasks. As we can use rules and goals to specify the characteristics of the corresponding relationships, it is not necessary to introduce additional design requirements.

# 5 Autonomy Model and Notation: Model

## 5.1 Meta Model Overview

Based on our 12 design requirements for the specification of autonomous agents and their interaction, we introduce a set of abstract constructs to develop a corresponding modeling language. We propose to name the modeling language *Autonomy Model and Notation* in the style of OMG's Business Process Model and Notation (BPMN) (Object Management Group 2013) or Decision Model and Notation (DMN) (Object Management Group 2016b) emphasizing that it consists of a general meta model and a specific graphical notation.

We are aware that our approach results in a complex meta model as well as notation. The intention is to provide a master blueprint to model autonomy. It may be to too complex to be used as a whole in simple cases or for specific applications. Hence, we propose to derive situational variants of the modeling method for local implementation. Approaches to do so, can be based on situational method engineering or configuration approaches (Delfmann 2006; Harmsen 1997; Janiesch 2007; Karlsson 2005).

We illustrate the resulting meta model in Figure 4. It incorporates the three categories *interfaces*, *nature*, and *behavior*. We can use the meta model to specify autonomy for smart objects in the IoT or CPS in general but also to describe virtual entities within (sub-)process in the domain of Business Process Management, or to characterize event processing agents and networks in Complex Event Processing. In general, we can use it to specify the degree of autonomy in procedural and analytical networks and to integrate the meta model with existing general-purpose modeling languages. Note that the meta model provide a blueprint of a super model. All aspects (i.e. interfaces, nature, and behavior) of the agent are optional so that a subset of the meta model can be used based on practical requirements.

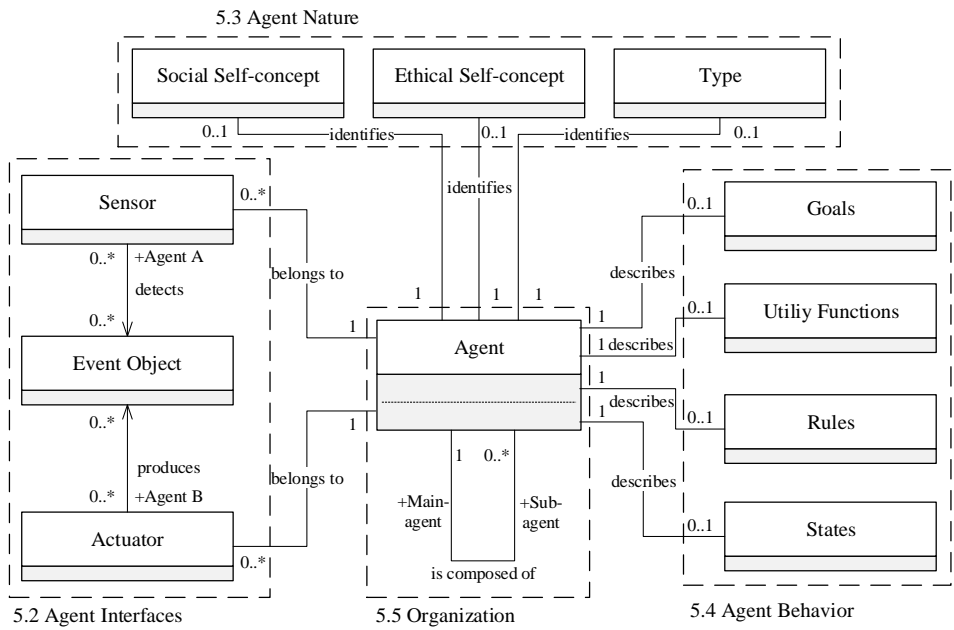

**Figure 4. AMN Meta Model**

## 5.2 Agent Interfaces

The interfaces category comprises an agent's sensors and actuators, which provide the technical foundation for event object recognition and distribution as specified by DR 2. Agents can be equipped with none or an arbitrary number of interfaces. In the traditional sense, each sensor belong to exactly one agent, artificial agents however can share sensors. The meta model incorporates this through the hierarchical *is composed of* relationship.

Two agents can interact, if their sensors and actuators are compatible, indicating that event objects generated by one agent can be recognized and processed by a different agent. We identify interfaces by certain types, which are based on the classic natural senses and include visual, auditory, olfactory, tactile, and gustatory modalities. In addition, we account for further senses of modern physiology (e.g., pain, temperature) and artificial transmission and recognition types with generic interfaces that are specified by passing additional parameters to cope with the entire range of event object types, for example electronic transmissions. An interaction relationship between two agents can be described further by the optional parameters attention, reliability, conformity, and security. See Figure 5 for an overview.

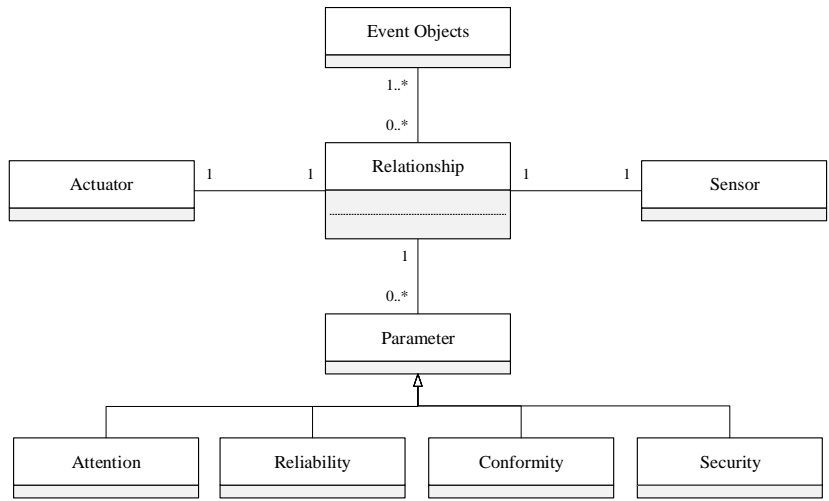

**Figure 5. Parameters of Agent Relationships**

To specify the capabilities of a sensor to recognize and process event objects, the concept of *attention* specifies a relationship's capacity for information processing (DR 6). Additionally, the constructs of *reliability* and *conformity* describe the degree of mutual trust between two agents (DR 8). Despite their similar structure, both concepts fundamentally differ in the direction of effect. *Reliability* defines the confidence of the event object receiver and, thus, indicates whether and to which degree the transferred information is considered during the decision-making process. By contrast, the construct of *conformity* describes the belief of the sender that the receiving agent transforms the exchanged information into conforming actions. Ultimately, we use the concept of *security* as an aggregate value to specify the freedom from corruption of the communication channel (cf. also DR 8). It could be extended to cover aspects of privacy, safety, or trustworthiness. As constituting characteristics of interactions between agents, we further distinguish between generic and specific event objects (DR 9, 10, and 12). Generic event objects have a flexible structure and composition and are used when no suitable specific event object is available. They can also be used to distinguish different types (autonomous) behavior for different application scenarios. For example, an autonomous electric car could distinguish the cases of driving, charging, and updating. Specific event objects include *reactions*, *tasks*, *action candidates*, *instructions*, *notifications*, and *metrics*. Reactions specify actions that are triggered by external requests or the sensor-based detection of an event. Although a reaction essentially describes a request-response-cycle, its potential variants, *acceptance*, *refusal*, and *veto*, are necessary to specify interaction patterns be-tween agents. A task defines a precise work assignment that entails agents defining and selecting adequate actions independently based on their specific goals. Action candidates specify potential actions and behaviors without the obligation for their implementation. By contrast, instructions require an agent's conformity when interpreting tasks and executing corresponding actions. Examples are process control instructions, such as instantiate, suspend, and abort. Ultimately, we use *notifications* to specify the exchanged information and *metrics* to report key performance indicators.

In addition to the different types of event objects, we also define two features that specify their overall structure. First, a *quantitative indicator* specifies the scope of an event object's content (DR 11). Thus, it can be reduced to a single aspect of the initial information, summarized to a preselection, or include all information. Second, the *media type* determines the configuration of an event object and is based on the previously described communication types and rules.

## 5.3 Agent Nature

The nature of an agent comprise its social and ethical self-concept as well as its type, which give an indication on how conflicts will be resolved between sub-agents or within a network and what its typical properties are (cf. DR 7). We specify an agent's social and ethical self-concept within our meta model to define its conduct within a group. Thus, we conceptualize an agent's capabilities to solve conflicts with other agents, for example between a super-agent and its sub-agents that result from differing or opposing goals, utility functions, rules, and states. See Figure 6 for the sets of social and ethical self-concepts. They can be used in any combination.

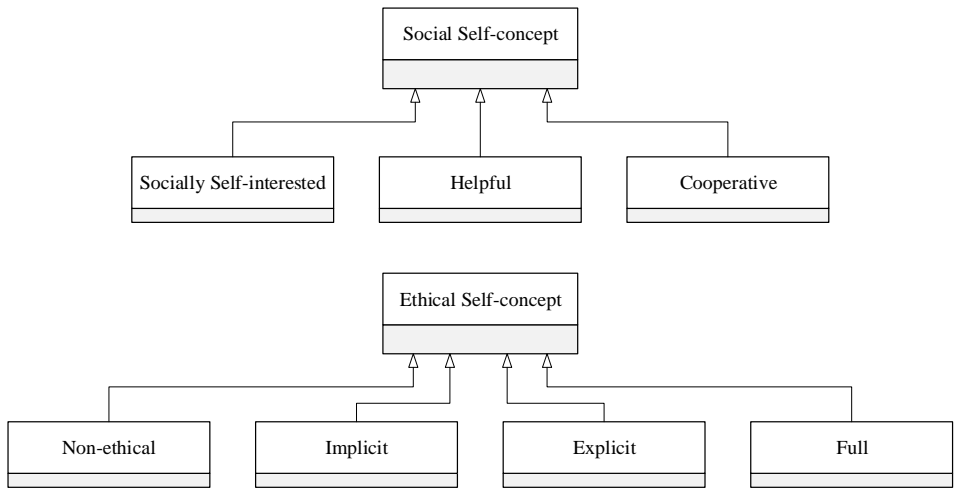

**Figure 6. Social and Ethical Self-concept and Ethics of Autonomous Agents**

*Socially self-interested* agents consider actions that maximize their individual utility, while accepting a comparably lower utility for the social group they are part of. By contrast, *helpful* agents generally aim for action candidates that influence the utility of their social group positively, although their individual benefits can be zero. Ultimately, *cooperative* agents collaborate with other agents to maximize the utility of the entire social group. Thus, they accept action candidates that are disadvantageous for them individually, as the resulting loss is compensated for by the benefits from collaboration.

*Non-ethical* agents do not consider ethics when maximizing their individual utility. This does not entail that they are unethical per se, they simply do not consider ethics. *Implicit ethical agents* have ethics built into their programming and the ethical responsibility is transferred to the developer. *Explicit ethical agents* use routines to evaluate action candidates based on their ethical standpoint. It is debatable as to whether machines can become *full ethical agents* rationalizing and justifying decisions using advanced AI technology or whether this category is reserved for humans.

While social and ethical skills account for an agent's specific autonomous behavior, we propose to further distinguish different agent types. While *functional* types provide means to model an agent's behavior in specific situations such as for analytical or procedural tasks, *autonomy types* specify high-level concepts that bundle functional types to distinguish agents on a more abstract level. Castelfranchi (2000) propose 13 types. For reasons of clarity, we condense them into nine types, which we summarize in Table 6. An agent can have one social self-concept and an agent can be of one type.

**Table 6. Types of Autonomous Agents (cf. Castelfranchi 2000)**

| Autonomy Type | Description |
| --- | --- |
| Interpretation | Agents capable of sourcing and processing relevant information independently (information, interpretation) |
| Know-how | Agents with sufficient cognitive capacity to achieve goals without instructions (know-how) |
| Plan | Agents capable of designing and applying plans to different conditions and possibly authorized to choose plans for other agents (planning, plan-discretion) |
| Goal | Agents striving to pursue and achieve their own goals, capable of choosing and prioritizing between multiple objectives and capable of and authorized to interrupt the execution of a plan and to adapt their preferences to changing conditions (motivational, goal-dynamics, goal-discretion) |

| Reasoning | Agents capable of drawing their own conclusion and acting accordingly (reasoning) |
|---|---|
| Monitoring | Agents capable of monitoring/checking their progress and success independently (monitoring) |
| Skill | Agents independent of other agents' skills to achieve a given goal (skill) |
| Resource | Agents independent of other agents' resources to achieve a given goal (resource) |
| Condition | Agents independent of other agents' approval or permission to achieve a given goal (condition) |

## 5.4 Agent Behavior

As the *behavior* of an agent, we specify agents' *goals*, *utility functions*, *rules*, and *states*, which are necessitated by DR 4 and 5. Their interplay describes the autonomous behavior of an agent. All of them are optional. First, goals constitute a list of objectives. As each goal determines an intended outcome, it serves as a criterion to evaluate the actions' quality. Since the environment is in constant flux, an agent must adapt its behavior and actions to unforeseen events and continuously changing structures. Thus, we use the concept of goals to account for an agent's adaptivity. Second, utility functions summarize an agent's preferences and allow the evaluation of multiple action candidates by determining their utility value. Third, rules provide means to define an agent's behavior explicitly based on multiple causal if-then-relationships. Closely linked to an agent's goals, rules provide a set of necessary actions to accomplish the predetermined objectives. Ultimately, states complete the set of describing characteristics by quantifying information about an agent that can have an influence on its actions and perceptions.

## 5.5 Agents Classes and Agent Structure

We can use two different approaches to structure and organize agents. Following DR 1, we need to organize agents hierarchically as part of a multi-agent system. As each agent can comprise multiple sub-agents, we can use a hierarchical representation to model agent networks on multiple levels with a varying information granularity. In addition, by connecting an arbitrary number of compatible sensors and actuators, we can loosely couple agents into a directed network, which satisfies DR 3.

To improve agent organization and model readability and clarity, we further provide concepts to aggregate agent instances to agent classes respectively, while accounting for instance individualities by cardinalities (cf. DR 1). Figure 7 illustrate the agent classes.

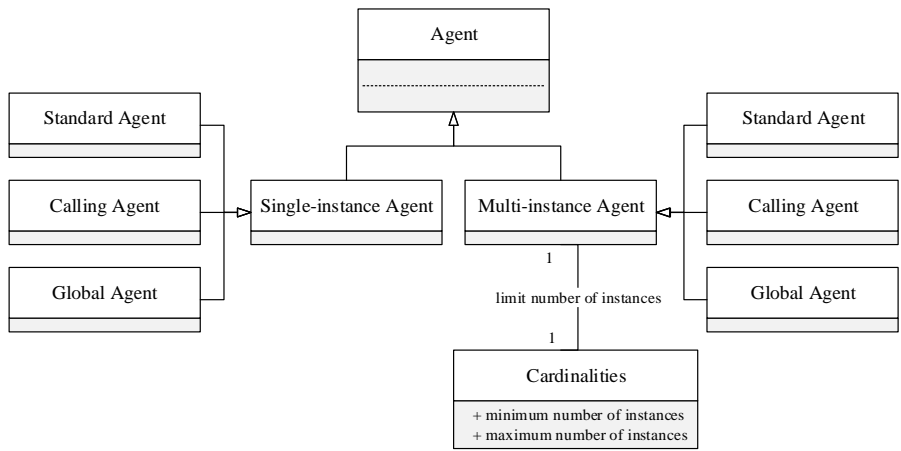

**Figure 7. Agent Classes and Instantiation**

We distinguish between the types of *single-instance* and *multi-instance agent*. We limit the number of possible instances by defining a range of minimum and maximum cardinalities for each multi-instance agent. We further specify these high-level agent types by the constructs of *standard agents*, *calling agents*, and *global agents*. Standard agents represent the general agent type as describe above. We use global agents as reusable building blocks. We further define calling agents, which participates in an agent network without the ability to establish hierarchical structures.

# 6  Autonomy Model and Notation: Notation

## 6.1  Notation Overview

Following the approach by Chen (2003), we transform our meta model into a graphical notation that is capable of capturing actions and behaviors of autonomous agents. We do this by analyzing the corresponding constructs semantically based on the principles of naturalness, resemblance, and differentiation. Second, we performed a literature analysis to collect design recommendations from established modeling languages to address possible user requirements and expectations. Third, we operationalize these findings to specify guidelines for the multi-level composition and grouping of constructs. Ultimately, we define a set of interpretation rules that facilitate the understandability of the resulting models. In addition, we draw upon the framework by Moody (2009), who provides recommendations for the design and visualizations of modeling constructs through nine principles. To further objectivize the development process, we also draw upon the recommendations of Schütte and Rotthowe (1998), who introduce the Guidelines of Modeling for the design of high-quality information models. Figure 8 summarizes the elements of our modeling notation.

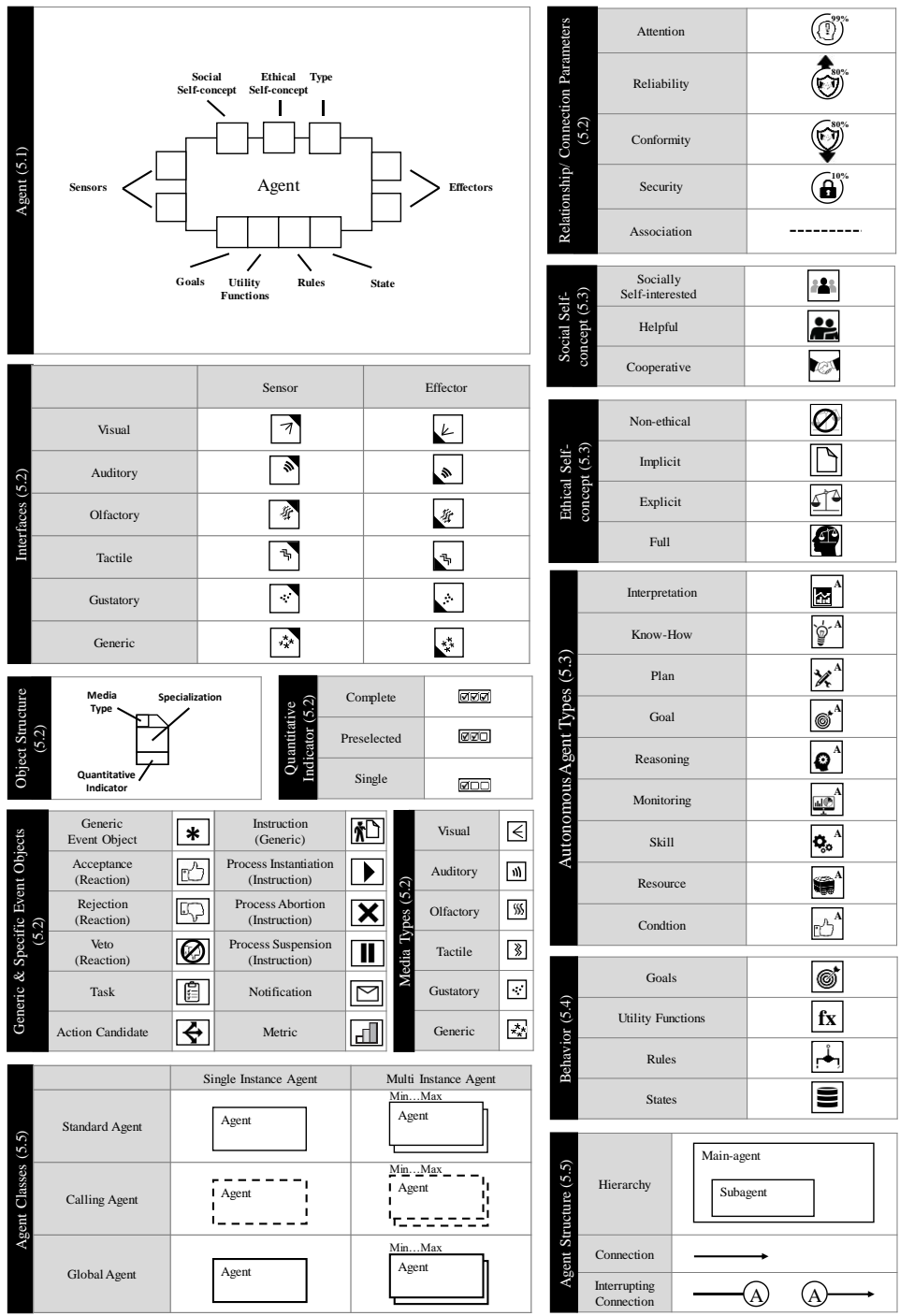

**Figure 8. AMN Notation**

## 6.2 Notation Details

**Agents (5.1):** As the most essential component of our modeling language, agents are represented by a rectangle. Based on the process of dual encoding, we can add a textual description to the graphical construct and, thus, ensure the unambiguous specification of different agents.

Alluding to the design of the Case Management Model and Notation (CMMN) language (Object Management Group 2016a), we can define an agent's characteristics by assigning graphical labels to predefined areas of the rectangle. As illustrated in Figure 8, these areas are represented by quadratic placeholders, which we specify subsequently. While labels differ in their function and icon design, we limit their shape to a square to account for the requirement of semiotic clarity. The areas left and right of the rectangle are dedicated to sensor and actuators constructs and specify an agent's communication capabilities. Labels for an agent's nature, that is its social self-concept and its type, are attached to the upper side of the rectangle. We further specify its behavior, including its goals, utility function, rules, and states in the lower side of the rectangle.

**Interfaces (5.2):** We describe the characteristics of an agent's interfaces by the five senses sight, hearing, taste, smell, and touch. To account for artificial communication forms, we add a generic sense that can be configured through additional parameters. We distinguish their specializations by integrating equal but mirrored icons for both interfaces. For added clarity, we color the upper right corner for sensors and the lower left corner for actuators respectively.

**Event object structure (incl. quantitative indicator and generic & specific event objects) (5.2):** We also provide structure for event objects exchanged between agents. They can be attached to the specifications of actuator-sensor connections between agents via associations. The blueprint for event objects comprises three areas that describe an object's media type, quantitative indicator, specialization. By combining these constructs, we can create detailed event object specifications. The media type results from the type of actuator that emitted the event object. We visualize an event object's media type by assigning the label of the corresponding interface to the upper left side of its basic structure. Second, the quantitative indicator describes whether the sending agent transfers all information or if for example action candidates have been preselected or narrowed down to a single option. We visualize this by adding checkboxes to the designated area of the object's structure. Ultimately, we specify the content of an event object by determining its specialization. As described in our meta model, we distinguish between generic and different specific event objects, ranging from reactions and tasks, to instructions, notifications, and metrics.

**Relationship/ connection parameters (5.2):** We further specify the characteristics of the connection, which is the relationship between agents, with the parameters attention, reliability, conformity, and security. As illustrated in Figure 8, all parameters are represented by a round symbol and augmented by a percentage value that quantifies the parameter's magnitude. Reliability describes the confidence from the perspective of an agent receiving an information object, while conformity specifies the confidence of the sender. Consequently, we use the same symbol for both constructs but account for their oppositional perspectives by adding directed arrows.

**Social self-concept (5.3):** As specified by our meta model, we use an agent's social self-concept to describe its strategy for solving conflicts in multi-agent networks. Thus, we distinguish between socially self-interested, helpful, and cooperative agents, which are all represented by squares with specific icons and appended to the upper side of an agent's basic structure.

**Ethical self-concept (5.3):** As specified by our meta model, we use an agent's ethical self-concept to describe its strategy for solving conflicts with regard to ethical problems. Thus, we distinguish between non-ethical, implicit, explicit, and full ethical agents, which are all represented by squares with specific icons and appended to the upper side of an agent's basic structure as well.

**Autonomous agent types (5.3):** We visualize the broad variety of autonomy types, which we append to the upper side of an agent's basic structure. To increase the visual distinguishability between both types, we further assign an "A" for autonomous agents and an "F" for functional agents to the upper right area of the corresponding symbol. At this stage, we only provide predefined autonomy types according to Table 6. An exemplary distinction of functional agents is the classification of Etzion and Niblett (2010) for analytical event processing agents in complex event processing networks.

**Behavior (5.4):** In addition, the agent's behavior includes goals, utility functions, rules, and states, which are modeled accordingly but appended to the lower side of the structure.

**Agent classes (5.5):** We describe agent classes in two dimensions. First, we distinguish between single instance and multi-instance agents, which we model either by using a regular rectangle or a stack of rectangles. While a single instance represents a unique agent, multi-instance agents summarize different agents with equal characteristics. We further highlight the minimum and maximum number of comprised instances by adding a textual annotation. Second, we capture the differentiation between standard, calling, and global agents. To improve model clarity, we vary the corresponding symbol's border. We draw standard agents with a regular border thickness. Global agents have a thicker border, while calling agents are also drawn with a thicker but dashed border.

**Agent structure (5.5):** We account for hierarchical structures within multi-agent systems by nesting multiple agent rectangles, so that main-agents can enclose an interdependent network of sub-agents. To visualize communication processes between agents on the same hierarchical level, we connect compatible sensors and actuators by a continuous or a discontinuous arrow with explicit connectors. We formally limit these constructs to single arrows to account for the unidirectional character of the communication process, which runs from actuators to sensors exclusively.

## 7 Illustration and Discussion

### 7.1 Scenario-based Illustration and Discussion

This situation is an excerpt and extrapolation from a scenario for CPS in smart factories provided by the German National Academy of Science and Engineering (acatech) (cf. Section 2.5.2 in Geisberger and Broy 2012). In the scenario, family Müller has placed an order with producer A to produce the furniture for their new kitchen. Producer A's smart factory can use multiple raw material to produce the desired furniture. Family Müller has selected a certain configuration of these. In the case of the scenario, a certain raw material becomes unavailable and the smart factory must use alternative materials.

In the following, we detail two options to deal with this issue. (1) As the result is equal in quality, the artificial production agent decides autonomously that a different material can be selected and gives the go ahead. (2) The artificial production agent does not have full autonomy and has to give family Müller a two-week window in which they can veto his decision to select a new material for production. He does not offer other options. This entails that the kitchen furniture will be delayed if the decision is vetoed. See Figure 9 for both variants (1) and (2) of the agent.

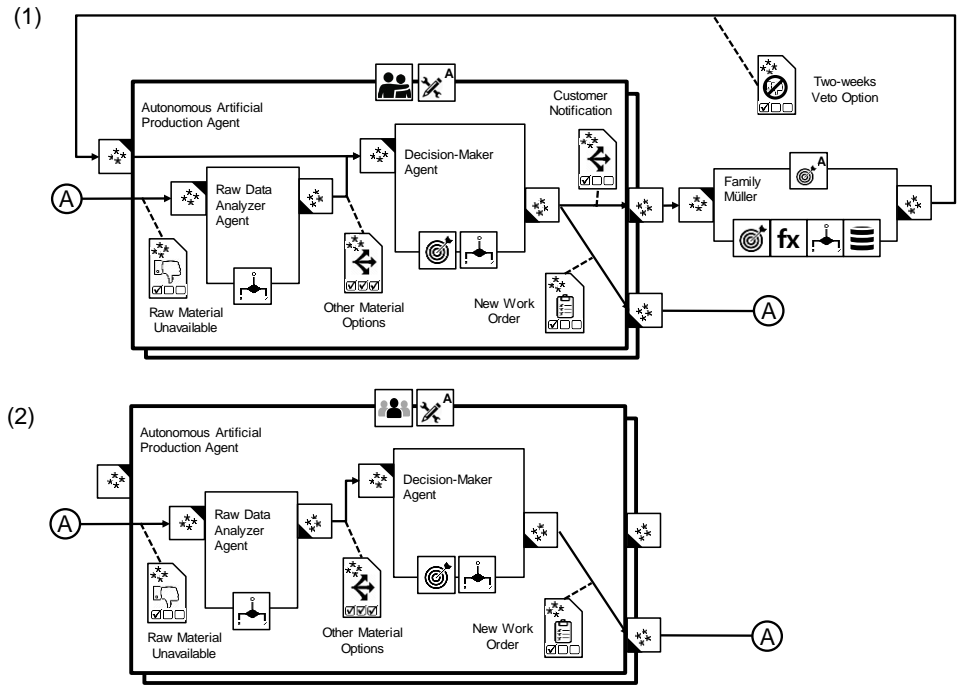

**Figure 9. Helpful Agent (A) vs. Socially Self-interested Agent (B)**

While this example is limited to a fraction of the possibilities to model autonomy with our notation, it highlights a couple of aspects that go beyond non-functional attributes and display the complexity of autonomous behavior.

In both cases, we have modeled the autonomous artificial production agent as a composition of a raw data analyzer agent, which is capable of determining raw material alternatives, and a decision maker agent, which has plan-discretion competence to issue a new work order. All communication takes place via e-mail or electronic data exchange. If a supplier (A) rejects or cancels a raw material order, the raw data analyzer agent will use his rules to provide the decision-maker agent with all other options to choose from. The decision-maker agent will then narrow down the selection to one action candidate according to his rules and goals.

In case (1) the agent is helpful, which entails that he will accept other materials that provide him zero direct benefit when the overall benefit (e.g., satisfied customer) is greater. He will prepare a work order with the new material of his choice and notify family Müller of his choice. If they do not veto within two weeks, he will communicate the decision to supplier (A). This behavior is consistent with autonomy level 6 of Table 5.

In case (2) the agent is socially self-interested and will select the most beneficial option according to his plan that is not overly detrimental to family Müller. He communicates his decision to supplier (A) and issues the task to produce with the new material of his choice. Family Müller will not even be notified. This is consistent behavior with autonomy level 1 of Table 5.

To specify other autonomous behavior, most autonomy levels could be employed and modeled to visualize the interaction between machine and human. We have modeled all levels of autonomy exemplarily and have not found a case we cannot support. In addition, other sensors and actuators could be used as well. Family Müller could for example be directly notified by a text to voice call on their phone.

In this example, we have not used relationship parameters to characterize the communication between the two agents. It is also conceivable that different communication channels could be used depending on their security needs or we could specify the relative reliability and conformity that a business assumes when communicating with their customers (i.e. in particular the trust that customers have in the decisions they make for them).

## 7.2 Criteria-based Discussion

To understand if our notation supports its users in the design of good conceptual models we evaluate whether the resulting models comply with the Guidelines of Modeling. We use the principles of Schütte and Rotthowe (1998) to do so: *construction adequacy*, *language adequacy*, *economic efficiency*, *clarity*, *systematic design*, and *comparability*.

While construction adequacy is satisfied if the behavior and structure of models are consistent with the real world, language adequacy demands compliance with the meta model's specifications. Building upon our scenario-based illustration, the resulting information models are consistent and complete against the meta model. We can transform all constructs from the meta model into notational elements and use them to specify their real-world relationships. Regarding construction adequacy, we modeled the behavior of different agent groups within the same multi-agent system and used the provided specifications to account for their individual structural requirements. Consequently, our notation facilitates its users to comply with these principles. Furthermore, our notation should not include constructs that are not relevant or eliminable without a considerable loss of meaning. To control for that, we continuously attempted to substitute or eliminate constructs during the modeling process in order to measure the importance and explanatory power of each notational element. However, we could not identify constructs that were frequently substituted or eliminated.

To provide adequate economic efficiency, Schütte and Rotthowe (1998) demand an advantageous cost-benefit-ratio during the model construction process. As autonomous multi-agent systems constitute a new modeling domain, corresponding conceptualizations of agents and event objects cannot be expressed through means other than informal diagrams or text. Due to the inherent abstraction through graphical visualizations, the models are easier to understand, quicker to access and asses than text-based or informal specifications, which positively influences modeling efficiency, effectivity, and productivity. Although we did not measure the necessary time and resources quantitatively, we observed it to be reasonably resource and time consuming. We further facilitate economic efficiency by reusing aggregated types of agents during the modeling process.

Although the scenario comprises a variety of agents and information objects, our notation remained clear and comparable. This is mostly due to the construct of interrupted connections that facilitates the decomposition of complex systems into smaller object systems, which can later be integrated by discontinuous actuator-sensor-connections. Textual annotations further allow specifying constitutional characteristics of agent systems, without increasing the overall complexity. We specified a set of syntactical and semantical modeling guidelines to ensure the comparability of the resulting information models. All models from our scenario are comparable, because of their equal structure and their compliance to these predetermined guidelines. The information models can be used for decision support, cross-system comparisons, and IT implementation analyses. We ensure systematic design of the resulting information models by providing comprehensive rules for connecting and integrating different perspectives, ranging from aggregated agents and agent instances to distributed multi-agent systems. We further provide interfaces for additional model layers that can enhance the current notation. For example, the rule construct facilitates the specification of methods and procedures for information analysis.

Moody (2009) proposes nine principles to develop a useful notation for conceptual models. He proposes the principles of *semiotic clarity* to limit symbol proliferation, *perceptual discriminability* to enable distinguishable symbols, *semantic transparency* to have the symbols' appearance suggest their meaning, *complexity management* to deal with inherent model complexity, *cognitive integration* to support model comparison and integration, *visual expressiveness* limiting visual variables, *dual coding* to supplement visual information, *graphic economy* to ensure cognitive processability, and *cognitive fit* to enable dialect for different audiences.

His principles of semiotic clarity postulates that symbols should not be redundant, overloaded, excessive or in deficit. We have abided to this principle by providing different simple icons as symbols for all necessary constructs. No construct is without symbol, symbols are not overloaded.

The principle of perceptual discriminability demands that symbols have a visual distance and avoid redundant coding. We have refrained from using colors as well as shapes at the same time to differentiate symbols. While interfaces, nature, and behavior of agents use rectangular symbols, parameters of the relationship have circular symbols. Event objects can be distinguished from agents by a dog-ear on the top-right corner.

To satisfy the principle of semantic transparency we have used icons as symbols in the notation that relate to results of Internet search for the corresponding keywords. We used Google's image search to do so.

The principle of visual expressiveness suggests limiting the visual vocabulary throughout the notation. We have refrained from using color as does UML for reasons of reproducibility. Further, we have limited the number of symbols (cf. autonomy types) and used graphical markers rather than textual annotations (cf. quantitative indicator).

In addition, we have followed the principle of dual coding to supplement visual information with annotations and through hybrid symbols for relationship parameters. We acknowledge that the principle of graphic economy, which ensures cognitive processability, may be violated for some users since we introduce more than 40 icons for our symbols. Yet, in a configurable modeling environment or when using pen and paper, symbols could always be balanced by text encoding for novice users. The latter also applies for the principle of cognitive fit.

The principles of complexity management and cognitive integration are related to the Guidelines of Modeling's principles of systematic design and have been discussed above.

In summary, we have created a graphical conceptual modeling language consisting of a meta model and a graphical notation that – in a first qualitative analysis – does neither violate the Guidelines of Modeling nor Moody's principles for visual notations but follows their recommendations wherever possible. It is conceivable to not only use the notation on its own but to integrate it with other notations to model the autonomy of business processes or process tasks in BPMN or (software) components in a UML diagram. This gives manifold access to model agents for smart cities, smart grids as well as smart mobility.

# 8 Summary, Limitations, and Conclusion

Agents can be specified with no regard for autonomy, with implicit autonomy, with explicit autonomy, or with human-like full autonomy. So far, the conceptual design of agents, which should act autonomously, has been predominantly implicit rather than explicit due to a lack of support in respective design languages. Based on a systematic literature review, we introduced 12 design requirements for the explicit specification of autonomous agents in the IoT and CPS. We used these design requirements to derive a meta model that allows for the explicit specification of autonomous agents. Based on the meta construct of interfaces, nature, and behavior, we can use the meta model to conceptualize agent systems that are further specified by sensors, actuators, exchanged event objects, social and ethical self-concept, type, goals, utility functions, rules, states, and their (hierarchical) composition. See Table 7 for the implementation of our design requirements.

Table 7. Implementation of Design Requirements in Conceptual Modeling Language

| DR | Design Requirement | Meta Model Entity |
|---|---|---|
| 1 | The language must support abstract and concrete hierarchical structures of autonomous agents. | Composition relation |
| 2 | The language must provide constructs to distinguish different types of sensors and actuators and specify which of them are supported by agents. Furthermore, it must specify the compatibility of these interfaces. | Sensor, actuators |

| 3 | The language must provide constructs that define an interaction's direction to reveal its inherent communication structure. | Types of compatible actuators and sensors |
|---|---|---|
| 4 | The language must provide a set of simple and complex rules to specify an agent's behavior and decision-making process. | Rules, actions |
| 5 | The language must provide constructs to specify different goals, utility functions, and states. | Goals, utility functions (preferences), states |
| 6 | The language must provide constructs that specify functional capacity restrictions, similar to the concept of attention. | Attention |
| 7 | The language must incorporate the concept of social and ethical rationali-ty. It must specify different types of social and ethical self-concepts and allow to group agents accordingly. This also applies to agents of the same type if their communication behavior is determined explicitly. Thus, the language must provide constructs to specify agent classes that differ in their number of potential and concrete instances. | Social self-concept (socially self-interested, helpful, cooperative), ethical self-concept (non-ethical, implicit, explicit, full), agent classes |
| 8 | The language must allow the specification of an agent's confidence in the communication channel and provide constructs for reliability, conformity, and security. | Reliability, conformity, security |
| 9 | The language must support task lists as a special type of data objects. | Event objects (generic/specific), media type |
| 10 | The language must provide data objects that indicate whether an agent accepts or declines a decision and orders to execute certain tasks. | Specific event objects |
| 11 | The language must provide data objects of the type task list corresponding to the scope of action candidates. | Event Object, quantitative indicator |
| 12 | The language must provide constructs to specify the type of a notification. | Specific event objects |

Based on the meta model, we developed a graphical notation that is augmented with syntactical and semantical guidelines. Together, both form the *Autonomy Model and Notation (AMN)*, a graphical conceptual modeling language.

We envision the modeling language to be combined with existing general-purpose and domain-specific modeling languages, which lack the aspect of autonomy. For example in UML, AMN could form an own perspective on components of software systems. Within BPMN, AMN could be used to describe the autonomous behavior pools of processes and their message flows.

We applied the modeling language to a smart factory scenario and illustrated as well as qualitatively evaluated its suitability and ability to support the construction of information models of autonomous agents.

We are taking a broad approach at modeling autonomy. Our approach is therefore applicable to multiple types of IoT applications, ranging from smart cars to smart factories to smart energy. As a consequence of this breadth, our modeling language may lack precision for some domains. There may be factors that influence autonomy in one case but do not in another. We do not cover these domain-specific factors. This is a limitation and a trade-off.

Approaches of this kind have a number of limitations. First, autonomous computing is continuously advancing in research and practice. Although our modeling language can account for a broad variety of requirements from the IoT and of current CPS, it may not yet be complete and it may demand periodic updates to ensure it is still current and applicable. Second, we must conduct more use cases to validate the suitability of our notation. Third, we must research whether our construct of an agent can be integrated with existing modeling languages and standards such as UML and BPMN to represent autonomy of agents in the IoT. In addition, larger and randomized quantitative analyses are necessary to assess the degree to which our modeling language is accepted by users. Recker et al. (2010) provide a baseline for the evaluation of conceptual models, which could be adapted.

Future research potentials especially result from the specification of autonomy interaction patterns between human and artificial agents according to the autonomy levels introduced. A modeling language also needs to be operationalized. Hence, we plan to develop a graphical modeling tool that supports the modeling process and increases the economic efficiency of our language. The advantages of the model-driven architecture deem that we research options to derive executable code or code skeletons from our conceptual models to further improve the economic efficiency of the implementation process. Ultimately, the model and notation could be extended to incorporate a more detailed security model incorporating enhanced aspects of privacy, safety, and trustworthiness. For example, interaction could be distinguished into at least anonymous, pseudonymized, and open. Trust is a wider issue that could be tackled in a distributed, possibly blockchain-based approach. If multiple agents report reliable and conforming behavior, agents should be more considered trustworthy.